\documentclass[showpacs,showkeys,twocolumn]{revtex4-1}
\usepackage{bm,graphicx,subfigure}
\begin{document}

\title{Vortex patterns in moderately rotating Bose-condensed gas}
\author{Mohd. Imran}\email{alimran5ab@gmail.com}
\author{M. A. H. Ahsan}
\affiliation{Department of Physics, Jamia Millia Islamia (Central University), New Delhi 110025, India.}
%\date{\today}

\begin{abstract}
Using exact diagonalization, we investigate the many-body ground state for vortex patterns in a rotating Bose-condensed gas of $N$ spinless particles, confined in a quasi-two-dimensional harmonic trap and interacting repulsively via finite-range Gaussian potential. The $N$-body Hamiltonian matrix is diagonalized in given subspaces of quantized total angular momentum $L_{z}$, to obtain the lowest-energy eigenstate. Further, the internal structure of these eigenstates is analyzed by calculating the corresponding conditional probability distribution. Specifically, the quantum mechanically stable as well as unstable states in a co-rotating frame are examined in the moderately rotating regime corresponding to angular momenta $4N \le L_{z} < 5N$ for $N=16$ bosons. In response to externally impressed rotation, patterns of singly quantized vortices are formed, shaping into canonical polygons with a central vortex at the trap center. The internal structure of unstable states reveals the mechanism of entry, nucleation and pattern formation of vortices with structural phase transition, as the condensate goes from one stable vortical state to the other. The stable polygonal vortex patterns having discrete $p$-fold rotational symmetry with $p=5$ and $p=6$ are observed. The hexagonal vortex pattern with $p=6$ symmetry is a precursor to the triangular vortex lattice of singly quantized vortices in the thermodynamic limit. For unstable states, quantum melting of vortex patterns due to uncertainty in positions of individual vortices, is also briefly discussed.
\pacs{05.30.Jp, 03.75.Hh, 03.75.Lm, 67.85.-d}
\keywords{Bose-Einstein condensation, Vortex patterns, Exact diagonalization, Finite-range Gaussian interaction potential, Conditional probability distribution.}
\end{abstract}
\maketitle

\section{Introduction}
\label{intro}
The experimental realization of Bose-Einstein condensate (BEC) with dilute vapors of ultra-cold alkali atoms in an external trap \cite{aem95,dma95,bst95} has become an important milestone in quantum many-body physics \cite{dgp99,ajl01,fs01,bpa10}.
These gaseous systems are dilute and inhomogeneous with controllable density, effective dimensionality and tunable atom-atom interaction of either sign \cite{ias98,cgj10}.
As a result of this experimental versatility, BEC has become an extremely convenient system to investigate macroscopic quantum phenomena such as superfluidity and quantum Hall physics \cite{ps02,jfa03,vie08}.
The formation of vortices with quantized circulation in response to rotation \cite{mcw00,hce01,hhh01,rav01,arv01,ech02,ech03,sce04} is intrinsically related to the existence of superfluidity.
Experimental efforts have further been focused on creating regular lattices with large number of singly quantized vortices \cite{rav01,arv01,ech02,ech03,sce04,ecs04}. 
On the theoretical front, studies such as in Refs.~\cite{br99,kmp00} have found successive transitions between stable patterns of singly quantized vortices.
The rotational properties of BEC and creation of vortices in a harmonic trap have been analyzed mostly by the mean-field approach like Gross-Pitaevskii scheme as in Refs.~\cite{br99,kmp00,lf99,lnf01,gp01,vvh05} or beyond the mean-field approximation \cite{kmp00,wg00,cwg01,pb01,ahs01,lhc01_pra,
ryl04,blo06,rkym06,ryb06,un06,dbo07,dbld09,bg10,ckm13}.
\\
\indent
In most of these studies, the two extreme regimes of a rotating BEC namely the slowly rotating regime and the rapidly rotating (quantum Hall) regime have been extensively explored, as summarized in several reviews \cite{vie08,bdz08,coo08,fet09,srh10}.
However, the study of regular patterns of few vortices in the intermediate regime of moderately rotating Bose-condensed gas has largely remained unexplored, more specifically using a full many-body approach such as exact diagonalization.
This regime is important from experimental view point also as the microscopic mechanism of entry, nucleation and formation of vortex patterns can be studied in a controlled fashion \cite{br99,cd99,kf04}.  
Further, the many-body correlation and quantum fluctuation play a significant role as the mean-field theory breaks down \cite{dbld09}.
A useful parameter controlling the degree of quantum fluctuation, namely, the filling fraction can conveniently be defined as $\nu=N/{N_{v}}$ where $N$ and $N_{v}$ are the number of bosons and the number of vortices, respectively \cite{cwg01}.
Quantum fluctuation is small for $\nu \rightarrow \infty$ but becomes increasingly significant  with decreasing value of $\nu$.
Sinova {\it et.\,al.}~\cite{shm02} have studied quantum fluctuation of vortex positions with decreasing $\nu$. 
An approximate value of $\nu$ where the mean-field theory breaks down specifically for the vortex lattice has been estimated \cite{cwg01}.
For $\nu > \nu_{c}$ the vortex lattice is stable where $\nu_{c} \sim 2-6$ from exact diagonalization calculation \cite{cwg01,cr07,lgv11} and $\nu_{c} \sim 8-14$ from the Lindemann criterion \cite{cwg01,shm02,gb04}.
\\
\indent
In this work, we present an exact diagonalization study of moderately rotating system of $N=16$ spinless bosons, interacting via short-range Gaussian repulsive potential in a quasi-two-dimensional harmonic trap.
Going beyond the slowly rotating regime, we focus our attention specifically on angular momentum subspaces $4N \le L_{z} < 5N$, well below the angular velocity for which the vortex lattice appears.
To obtain the $N$-body lowest-energy eigenstate corresponding to stable and unstable states in the co-rotating frame, exact diagonalization of the many-body Hamiltonian matrix is carried out using Davidson iterative algorithm \cite{dav75} in given subspaces of quantized total angular momentum $L_{z}$.
The aim of the present work is to analyze in the moderately rotating limit the quantum mechanically stable as well as unstable states and their internal structure (spatial correlation) by calculating the conditional probability distribution of the corresponding eigenstates \cite{ckm13}.
Our analysis is based on the premise that the formation of stable vortex patterns with definite discrete rotational symmetry and its structural phase transition under rotation from one stable state to the other, can be understood in terms of quantum fluctuation (leading to quantum melting) due to uncertainty in positions of individual vortices in the intervening unstable states.   
\\
\indent
This paper is organized as follows.
In Sec.~\ref{model}, we describe the model Hamiltonian for the rotating Bose gas interacting via finite-range Gaussian potential.
Subsequently the single-particle reduced density matrix is introduced to delineate the macroscopic condensate and its  vorticity.
A brief description of conditional probability distribution as a measure of internal structure of the many-body eigenstates is presented next. 
In Sec.~\ref{results}, we present our results on a moderately rotating system of bosons to analyze the internal structure of quantum mechanically stable as well as unstable states in the co-rotating frame. 
Finally in Sec.~\ref{conc}, we summarize our main results.

\section{The Model}
\label{model}
We consider a system of $N$ interacting spinless bosons, harmonically confined and subjected to an externally impressed rotation about the $z$-axis at an angular velocity ${\widetilde{\bm{\Omega}}}\equiv \widetilde{\Omega} \hat{e}_{z}$.
We assume a stiff confinement of the external trap potential $V\left({\bf r}\right)=\frac{1}{2}M\left({\omega}^{2}_{\perp}r^{2}_{\perp}+{\omega}^{2}_{z}z^{2}\right)$ along the axis of rotation 
{\it i.e.} $\omega_{z} \gg \omega_{\perp}$ so that the many-body dynamics along the $z$-axis is frozen, yielding an effectively quasi-2D system with $x$-$y$ rotational symmetry.
Here, $r_{\perp}=\sqrt{x^{2}+y^{2}}$ is the radial distance from the trap center, $M$ is the mass of an atom, $\omega_{\perp}$ and $\omega_{z}$ are the radial and the axial frequencies, respectively, of the harmonic confinement.
We chose $\hbar \omega_{\perp}$ as the unit of energy and $a_{\perp} = \sqrt{\hbar/{M \omega_{\perp}}}$ as the corresponding unit length.
Introducing $\Omega \equiv \widetilde{\Omega}/{\omega_{\perp}}$ $(\leq 1)$ as the dimensionless angular velocity and $L_{z}$ (scaled by $\hbar$) being the $z$ projection of the total angular momentum operator, the many-body Hamiltonian in the co-rotating frame is given by ${H}^{rot}={H}^{lab}-\Omega L_{z}$ where
\begin{equation}
H^{lab} = \sum_{j=1}^{N} \left[-\frac{1}{2} \bm{\nabla}^{2}_{j} + \frac{1}{2} {\bf r}_{j}^{2} \right] + \frac{1}{2} \sum_{i\neq j}^{N} U \left({\bf r}_{i},{\bf r}_{j}\right).
\label{mbh}
\end{equation} 
The first two terms in the Hamiltonian~(\ref{mbh}) correspond to the kinetic and potential energies respectively.
The third term $U\left({\bf r},{\bf r}^{\prime}\right)$ arises from the two-body interaction assumed to be Gaussian in particle-particle separation~\cite{iasl15}
\begin{equation}
U \left({\bf r},{\bf r}^{\prime}\right) = \frac{\mbox{g}_{2}} {2\pi{\sigma^{2}}}
\exp{\left[ -\frac{\left(r_{\perp}-r_{\perp}^{\prime}\right)^{2}}{2\sigma^{2}} \right]} 
\delta \left(z-z^{\prime}\right)
\label{gip}
\end{equation}
with $\sigma$ (scaled by $a_{\perp}$) being the effective range of the Gaussian potential.
The dimensionless parameter $\mbox{g}_{2}=4\pi {a_{s}}/{a_{\perp}}$ is a measure of the strength of interaction where $a_{s}$ is the $s$-wave scattering length for low-energy particle-particle collision.
In view of the recent advancements in atomic physics, it has become possible to tune the low-energy atom-atom scattering length in ultra-cold atomic vapors using Feshbach resonance \cite{ias98,cgj10}.
In the present work, the scattering length is taken to be positive $\left(a_{s}>0\right)$ so that the effective interaction is repulsive.
In addition to being physically more realistic, the finite-range Gaussian interaction potential~(\ref{gip}) is expandable within a finite number of single-particle basis functions and hence computationally more feasible compared to the zero-range $\delta$-function potential \cite{cfa09,dka13}.  
In the limit $ \sigma \rightarrow 0$, the normalized Gaussian potential in Eq.~(\ref{gip}) reduces to the zero-range contact potential $\mbox{g}_{2}\delta\left({\bf r}-{\bf r}^{\prime}\right)$, which has been used in earlier studies \cite{dgp99}. 
\\
\indent
The system described by the Hamiltonian in Eq.~(\ref{mbh}) has cylindrical symmetry where the $z$-projection of total angular momentum is conserved {\it i.e.} $L_{z}$ is a good quantum number.
To obtain the many-body  eigenstates, we employ exact diagonalization of the Hamiltonian matrix in different subspaces of $L_{z}$ with inclusion of lowest as well as higher Landau levels in constructing the $N$-body basis states \cite{ahs01,lhc01_pra}.
The Hamiltonian $\hat{H}^{lab}$ in Eq.~(\ref{mbh}) is diagonalized in given subspaces of $L_{z}$ to obtain the energy $E^{rot}\left(L_{z},\Omega\right)=E^{lab}\left(L_{z}\right)-\Omega L_{z}$ in the co-rotating frame.  
This is equivalent to minimizing $E^{lab}\left( L_{z}\right)$ subject to the constraint that the system has angular momentum expectation value $L_{z}$ with angular velocity $\Omega$ identified as the corresponding Lagrange multiplier.
Fixing $L_{z}$, therefore, fixes $\Omega$ and accordingly we mention $L_{z}$ (instead of $\Omega$), in all tables and figures throughout this work.
\paragraph*{Single-particle reduced density matrix (SPRDM).}
The $N$-body ground state wavefunction $\Psi_{0}\left({\bf r}_{1},{\bf r}_{2},\dots,{\bf r}_{N}\right)$ is assumed to be normalized; one can then determine the single-particle reduced density matrix $\rho \left({\bf r},{\bf r}^{\prime}\right)$, by integrating out the degrees of freedom of $\left(N-1\right)$ particles.
Thus
\begin{eqnarray}
\rho \left({\bf r},{\bf r}^{\prime}\right) &=&\int \int \dots \int d{\bf r}_{2}\ d{\bf r}_{3}\dots d{\bf r}_{{N}} \nonumber \\
&&\times\ \Psi_{0}^{\ast}({\bf r},{\bf r}_{2},{\bf r}_{3} \dots,{\bf r}_{{N}})\ \Psi_{0}({\bf r}^{\prime},{\bf r}_{2},{\bf r}_{3},\dots,{\bf r}_{{N}}) \nonumber \\
&\equiv & \sum_{{\bf n},{\bf n}^{\prime}}\ \rho_{_{{\bf n},{\bf n}^{\prime}}}\ u^{\ast}_{\bf n}\left({\bf r}\right)u_{{\bf n}^{\prime }}\left({\bf r}^{\prime }\right).  
\end{eqnarray}
The above expression is written in terms of single-particle basis functions $u_{\bf n}\left({\bf r}\right)$ with quantum number ${\bf n}\equiv \left(n,m\right)$.
Being hermitian, this can be diagonalized to give 
\begin{equation}
\rho \left({\bf r},{\bf r}^{\prime}\right) = \sum_{\mu }\lambda_{\mu } \ \chi^{\ast}_{\mu }\left({\bf r}\right)
\chi_{\mu } \left({\bf r}^{\prime }\right),
\label{spd}
\end{equation} 
where $
\chi_{\mu }\left({\bf r}\right) \equiv \sum_{\bf n} c^{\mu }_{\bf n} \, {u}_{\bf n}\left({\bf r}\right)$ and $\sum_{\mu }\lambda_{\mu }=1$
with $ 1\geq\lambda_{1}\geq\lambda_{2}\geq\cdots\lambda_{\mu }\geq\cdots \geq 0 $.
The $\left\{\lambda_{\mu }\right\}$ are the eigenvalues, ordered as above, and $\left\{ \chi_{\mu }\left({\bf r}\right)\right\}$ are the corresponding eigenvectors of the SPRDM (\ref{spd}); each $\mu$ defines a fraction of the BEC.
Thus, for a particular $L_{z}$-state, every fraction $\lambda_{\mu}$ of the SPRDM is characterized by a unique value of single-particle angular momentum quantum number $m_{\mu}$.
For such a system, the {\it vorticity} is identified by the angular momentum quantum number $m_{1}$ of the most dominant single-particle state $\chi_{1}\left({\bf r}\right)$ corresponding to the largest eigenvalue $\lambda_{1}$ of the SPRDM.
\paragraph*{Conditional probability distribution (CPD).}
The internal structure (spatial correlation) of a many-body state can be analyzed by calculating the conditional probability distribution (CPD) \cite{blo06,ckm13,yl00} defined as
\begin{equation}
\mathcal{P} \left({\bf r},{\bf r}_{0}\right) = \frac{\langle \Psi\vert \sum_{i \neq j} \delta \left({\bf r}- {\bf r}_{i} \right)\delta \left({\bf r}_{0}-{\bf r}_{j}\right) \vert \Psi \rangle}{\left(N-1\right)\sum_{j} \langle \Psi\vert \delta \left({\bf r}_{0}- {\bf r}_{j} \right)\vert \Psi \rangle}
\label{cpd}
\end{equation}
where $\vert \Psi \rangle$ is the many-body ground state obtained through exact diagonalization and ${\bf r}_{0}=(x_{0},y_{0})$ is the reference point (usually chosen to be the position of high density for a few-body system like ours).
The CPD can be interpreted as the probability of a particle being at position ${\bf r}$ under the condition that another one is located (fixed) at ${\bf r}_{0}$ \cite{lhc01_pra}.
The calculation of CPD for many-body states with different values of $L_{z}$ provides information about the pattern of vortices, discrete $p$-fold rotational symmetry and size of the condensate in a harmonic trap (for instance, see Fig.~\ref{fig:ltc}).

\section{Results and Discussion}
\label{results} 
\begin{figure*}[!htb]
\subfigure[$\,L_{z}=69$]{\includegraphics[width=0.19\linewidth]{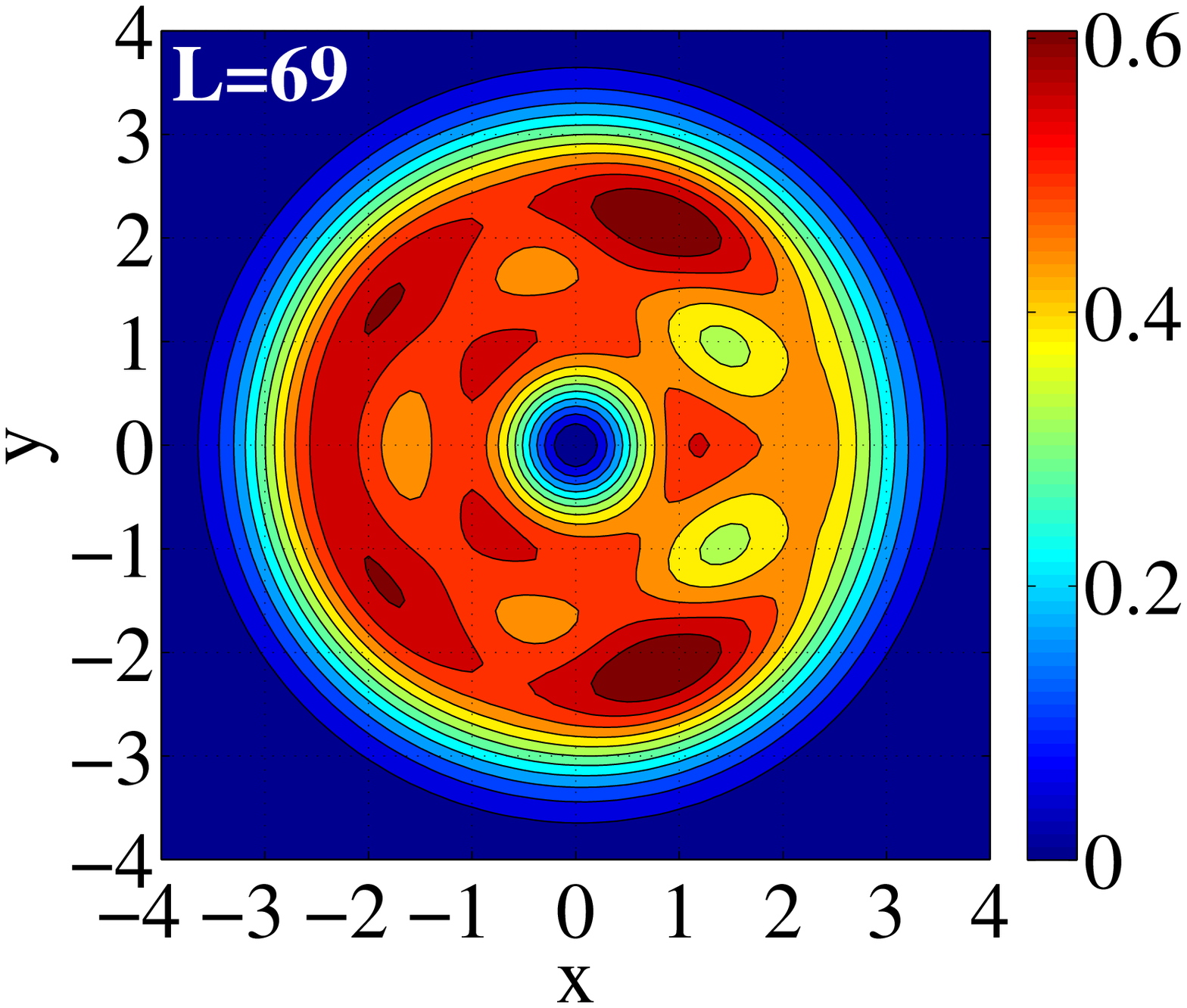}\label{fig:ltc69}}
\subfigure[$\,L_{z}=70$]{\includegraphics[width=0.19\linewidth]{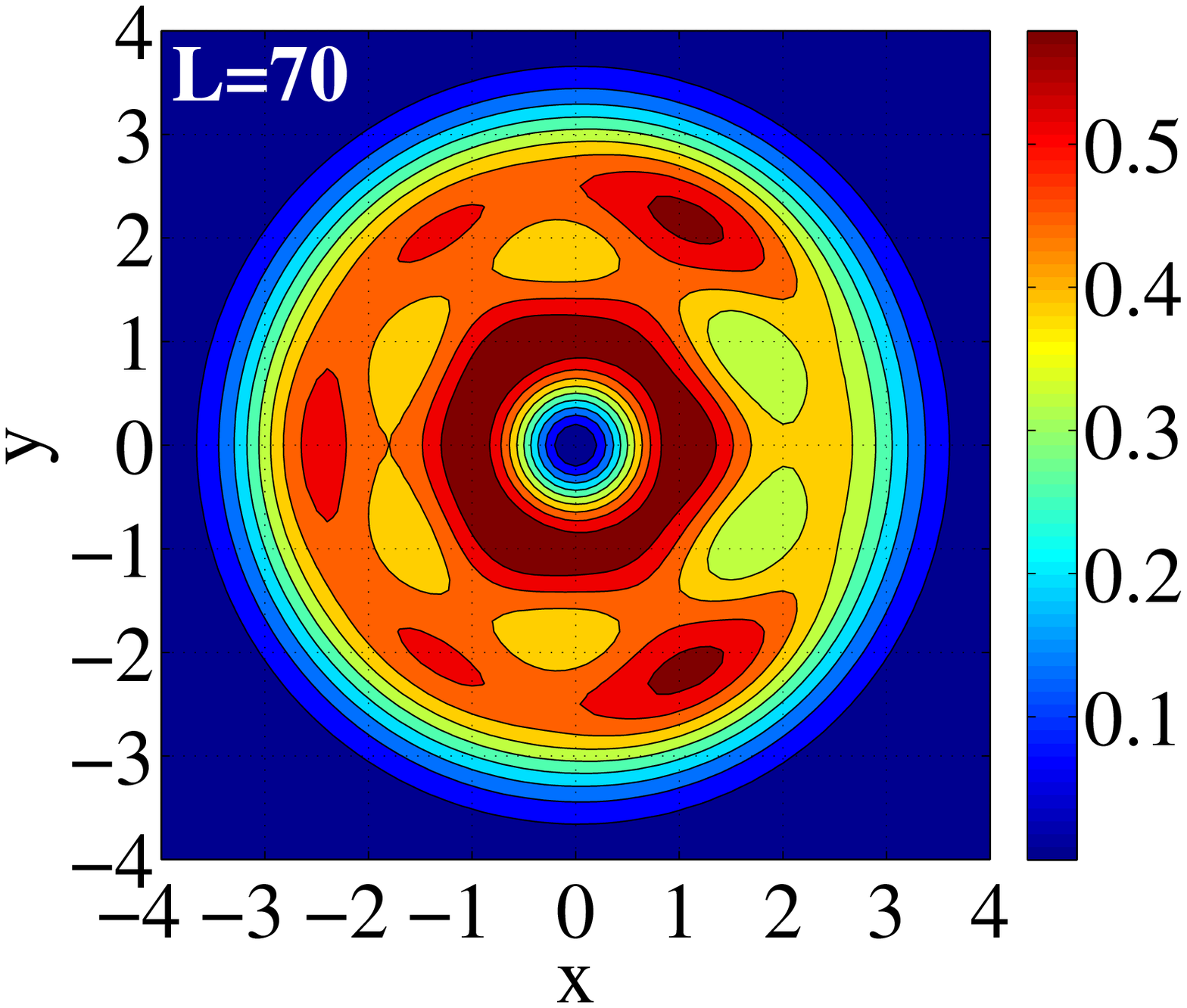}\label{fig:ltc70}}
\subfigure[~Stable $\,L_{z}=71$]{\includegraphics[width=0.19\linewidth]{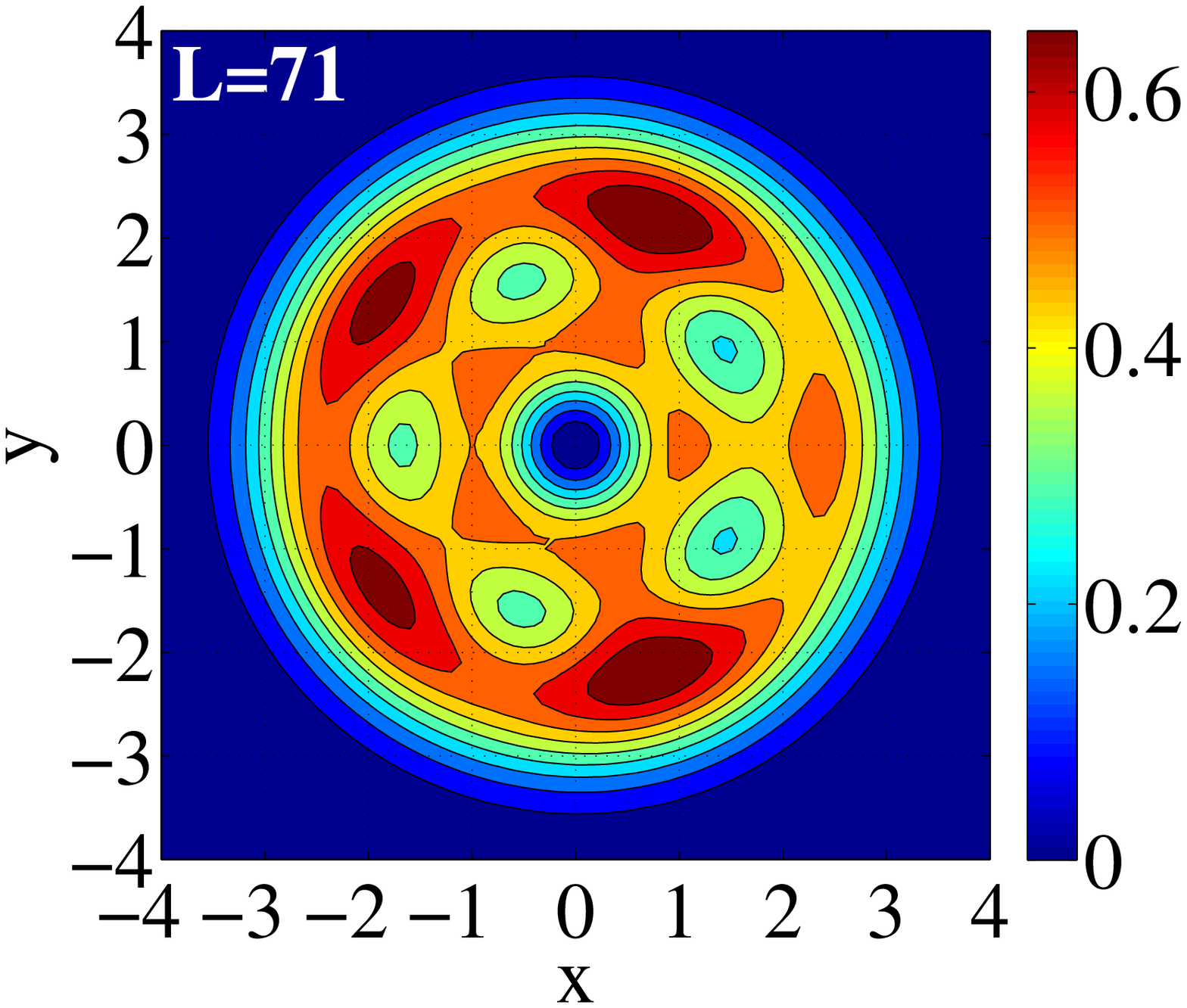}\label{fig:ltc71}}
\subfigure[$\,L_{z}=72$]{\includegraphics[width=0.19\linewidth]{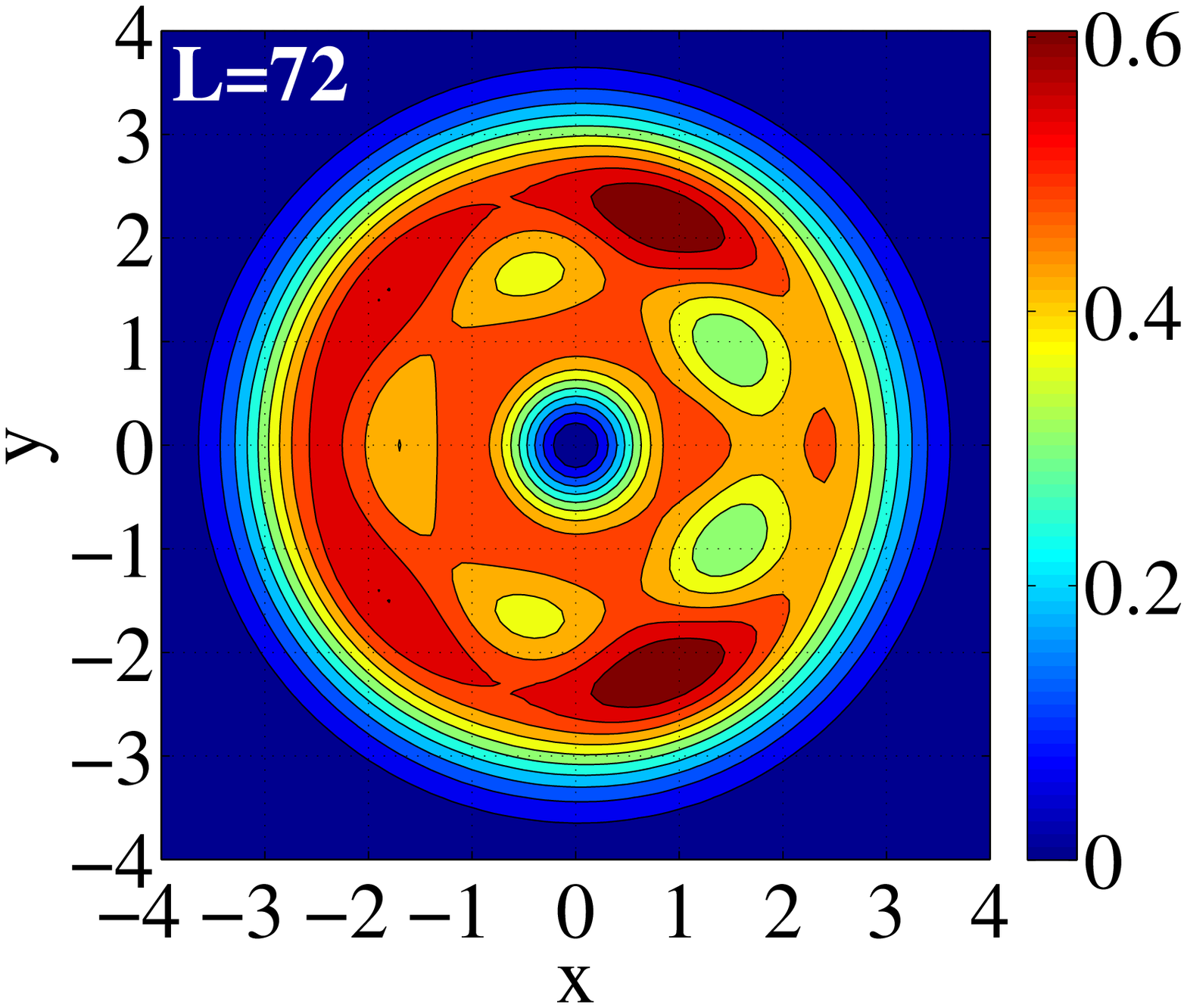}\label{fig:ltc72}}
\subfigure[$\,L_{z}=73$]{\includegraphics[width=0.19\linewidth]{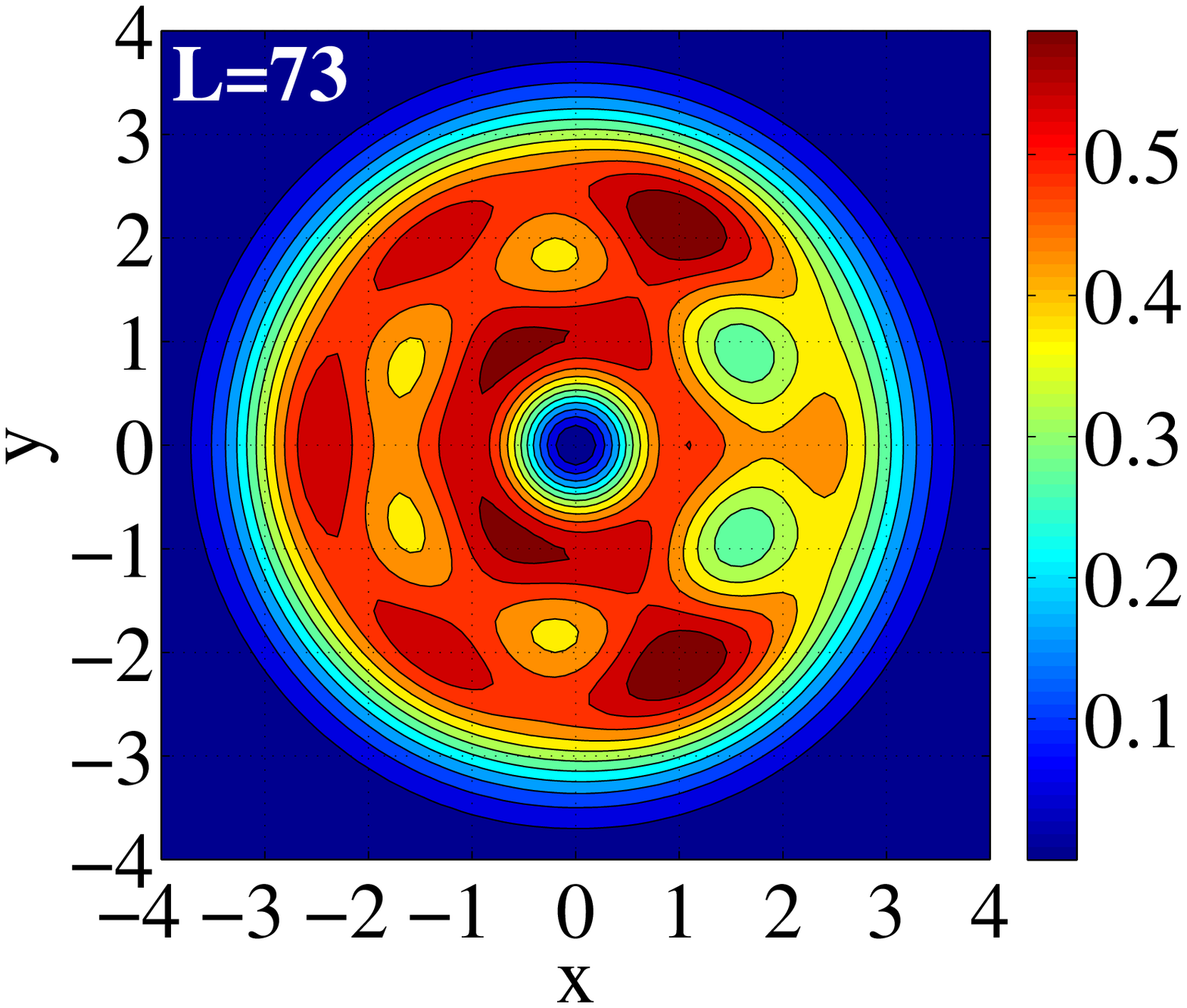}\label{fig:ltc73}}
\subfigure[$\,L_{z}=74$]{\includegraphics[width=0.19\linewidth]{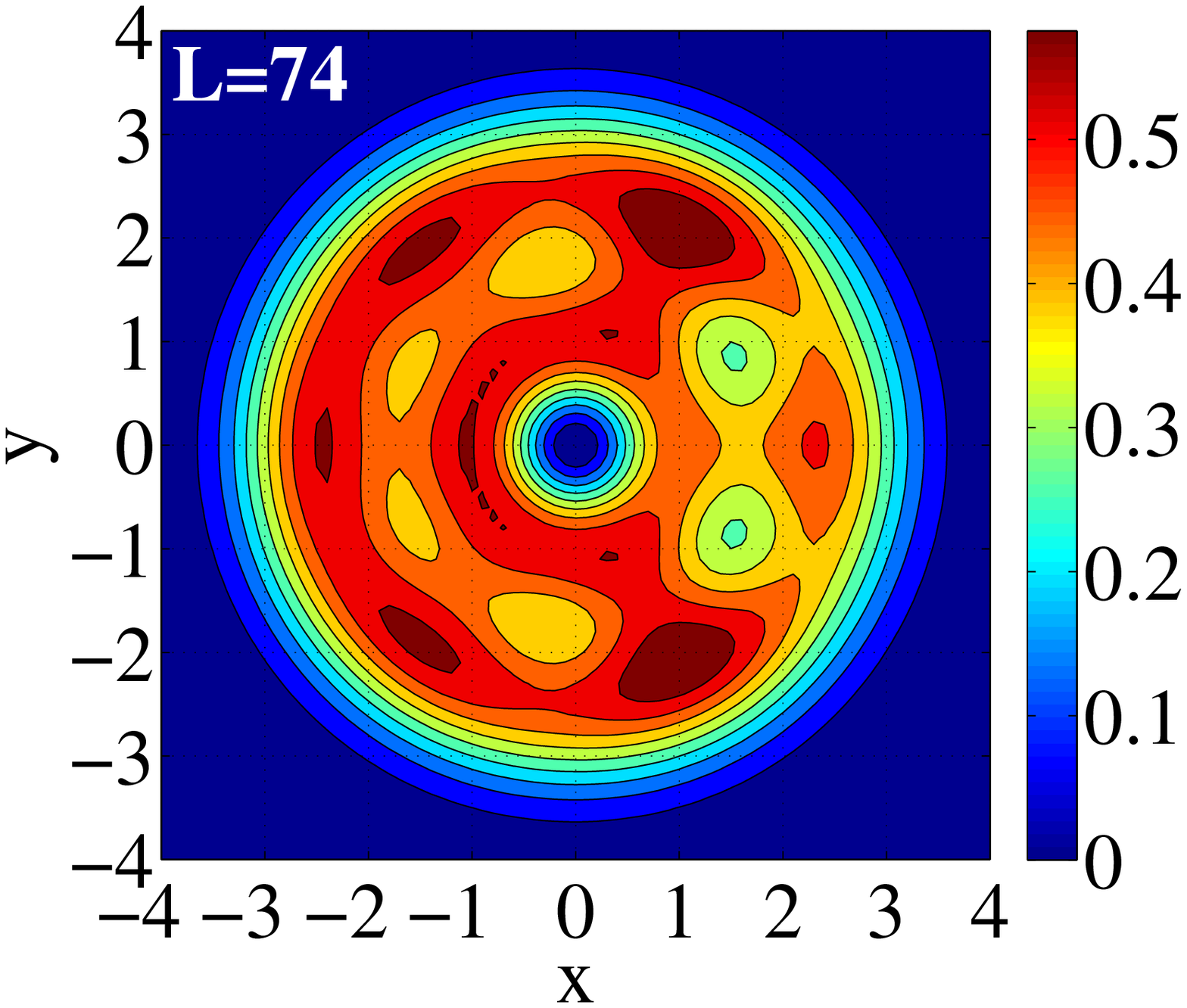}\label{fig:ltc74}}
\subfigure[$\,L_{z}=75$]{\includegraphics[width=0.19\linewidth]{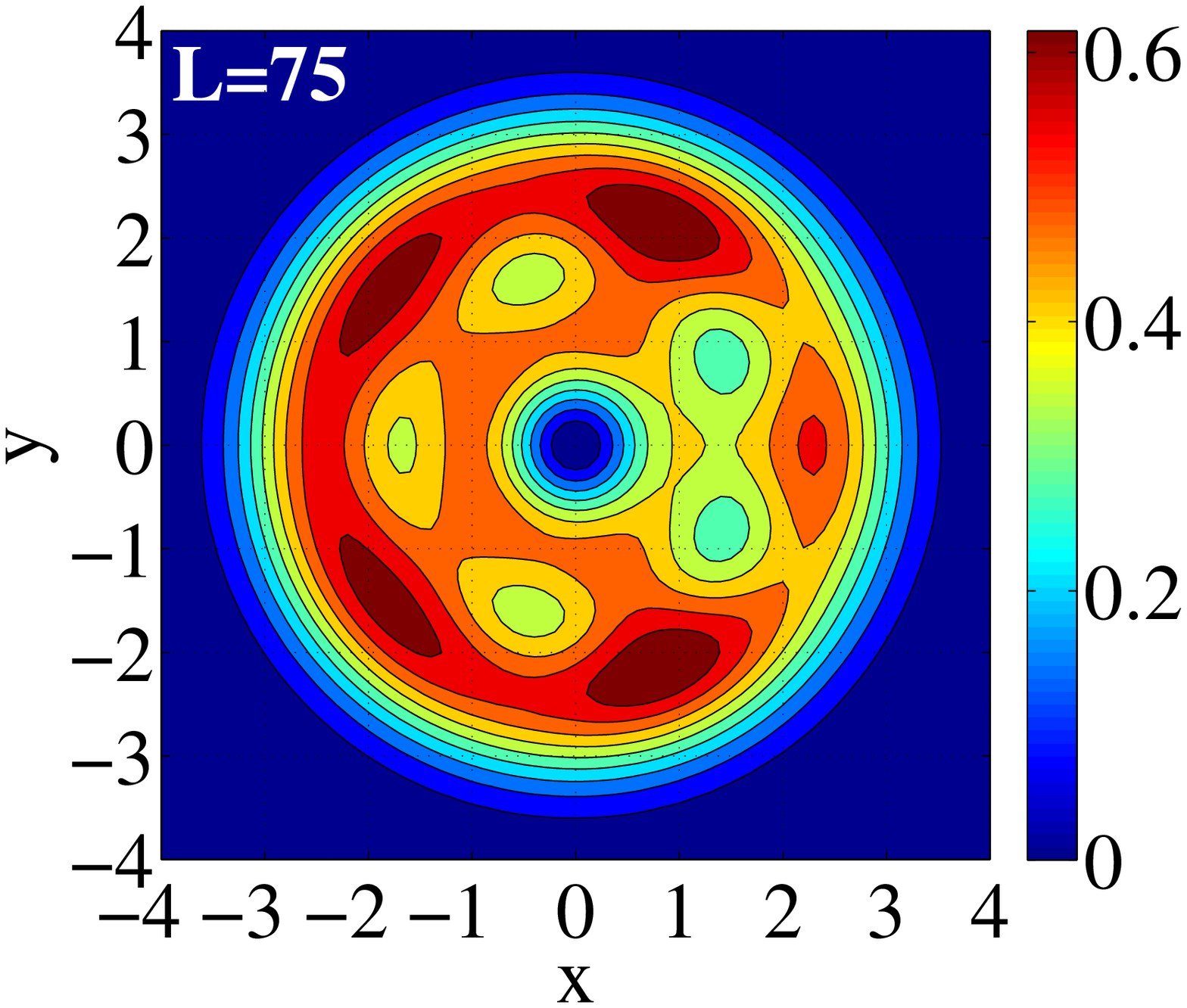}\label{fig:ltc75}}
\subfigure[~Stable $\,L_{z}=76$]{\includegraphics[width=0.19\linewidth]{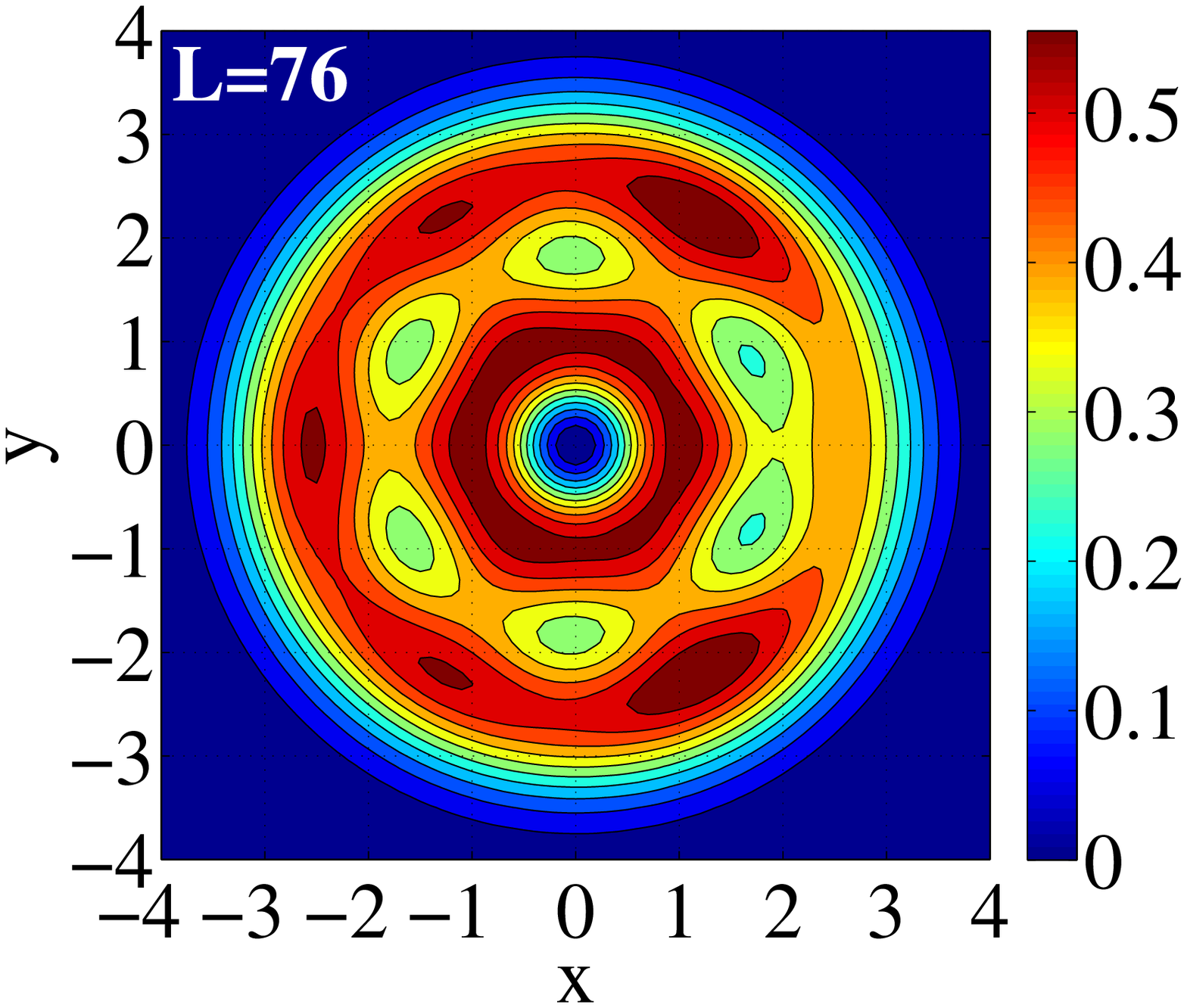}\label{fig:ltc76}}
\subfigure[$\,L_{z}=77$]{\includegraphics[width=0.19\linewidth]{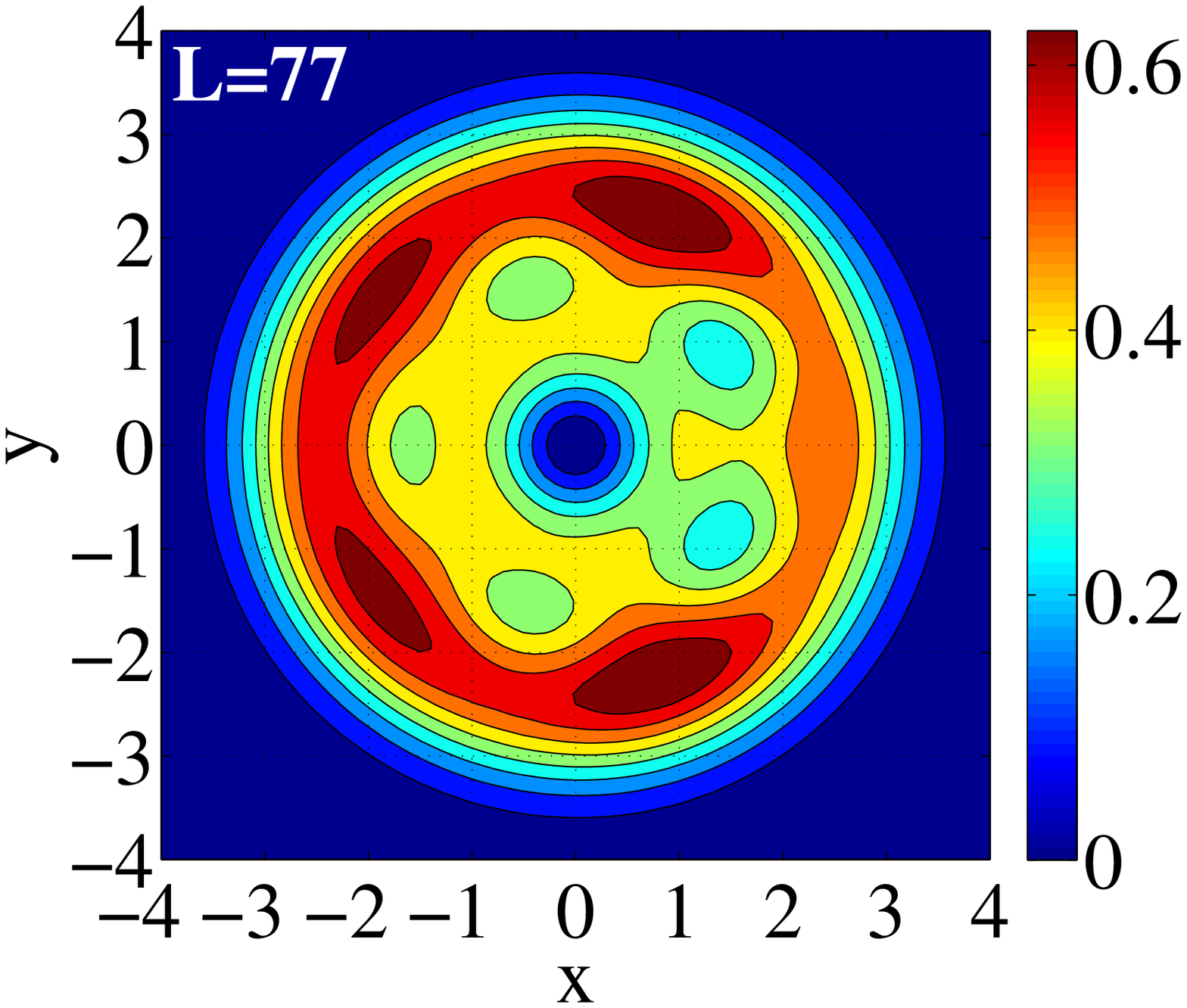}\label{fig:ltc77}}
\subfigure[$\,L_{z}=78$]{\includegraphics[width=0.19\linewidth]{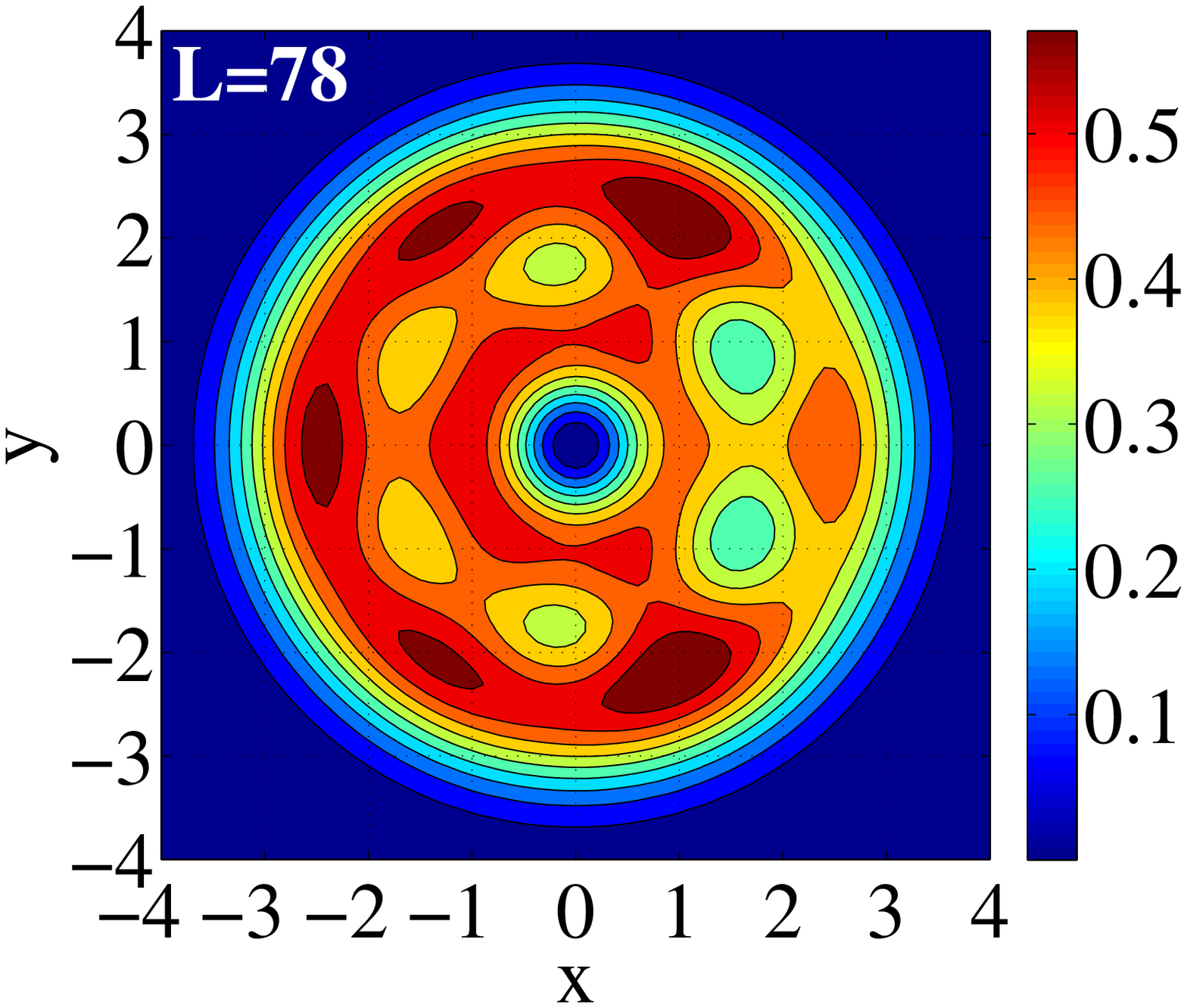}\label{fig:ltc78}}
\caption{\label{fig:ltc}(Color online) CPD contour plots of angular momentum states $L_{z}=69$ to $78$ for $N=16$ bosons with interaction parameters $\mbox{g}_{2}=0.9151$ and $\sigma=0.1$ in the Gaussian potential~(\ref{gip}). The angular momentum states $L_{z}=71\ \mbox{and}\ 76$ are the stable ones and the rest of the states are unstable in the co-rotating frame. Each contour plot is an isosurface density profile viewed along $z$-axis, the axis of rotation. The reference point for CPD plots is located at ${\bf r}_{0} = \left(x_{0}, y_{0}\right) = \left(1.5, 0\right)$ in units of $a_{\perp}$. Brown-red region has the highest probability density falling off to blue region of lowest probability density, as shown on the color bar.}
\end{figure*}

The results presented here are for a rotating system of $N=16$ Bose atoms of $^{87}$Rb confined in a  quasi-2D harmonic trap, with radial frequency $\omega_{\perp}=2\pi \times 220$ Hz and aspect ratio of $\lambda_{z} \equiv \omega_{z}/ {\omega_{\perp}}=\sqrt{8}$. 
This choice of radial frequency corresponds to the trap length $a_{\perp}=\sqrt{{\hbar}/{M\omega_{\perp}}}=0.727 \mu m$.
The condensate has extension $a_{z}=\sqrt{\hbar/M\omega_{z}}=a_{\perp}\lambda_{z}^{-{1}/{2}}$ in the $z$-direction and its dynamics along this axis is assumed to be completely frozen.
It is to be noted that for a many-body system (under consideration here), the characteristic energy scale for the interaction is determined by the dimensionless parameter $\left(Na_{s}/a_{\perp}\right)$. 
Owing to the increasing dimensionality of the Hilbert space with $N$, the computation soon becomes impractical.
Therefore, we vary $a_{s}$ to achieve a suitable value of  $\left(N a_{s}/a_{\perp}\right)$ relevant to experimental situation \cite{dgp99}.
The parameters of Gaussian interaction potential in Eq.~(\ref{gip}) have been chosen as follows: range $\sigma = 0.1$ (in units of $a_{\perp}$) and $s$-wave scattering length $a_{s}=1000a_{0}$, where $a_{0}=0.05292~nm$ is the Bohr radius. 
The corresponding value of the dimensionless interaction parameter $\mbox{g}_{2}$ turns out to be $0.9151$ leading to $(Na_{s}/a_{\perp}) \sim 1$ in the moderately interacting regime \cite{ahs01,lhc01_pra}.
Further, the system is subjected to an externally impressed rotation along $z$-axis with dimensionless angular velocity $\Omega \equiv {\widetilde{\Omega}}/{\omega_{\perp}}$.
The simultaneous eigenstate of Hamiltonian and total angular momentum minimizes the free energy at zero-temperature in the co-rotating frame to become the ground state of the system.
With the usual identification of $\Omega$ as the Lagrange multiplier associated with the total angular momentum $L_{z}$ for the rotating system, the $L_{z}$-$\Omega$ stability line has a series of critical angular velocities $\Omega_{{\bf c}i},\ i=1,2,3,\cdots$, at which total angular momentum of the condensed many-body ground state takes quantum jump (undergoes quantum phase transition) \cite{ahs01}.
The ground state corresponding to critical angular velocity $\Omega_{{\bf c}i}$, is referred to as quantum mechanically stable phase-coherent vortical state \cite{lf99,lnf01,gp01}.
\\
\indent
The many-body ground state wavefunction, in the beyond lowest Landau level approximation \cite{ahs01,lhc01_pra}, is obtained through exact diagonalization of the Hamiltonian matrix using Davidson iterative algorithm \cite{dav75}.  
The diagonalization is carried out separately for each of the subspaces of quantized total angular momentum $L_{z}$ and the ground state energy $E^{rot}\left(L_{z},\Omega\right)=E^{lab}\left(L_{z}\right)-\Omega L_{z}$ of the rotating condensate calculated in the co-rotating frame. 
For a rotating Bose-condensed gas with total angular momentum $L_{z}=N$, a single vortex aligned with the trap center appears \cite{mcw00,br99}. 
As the angular velocity is increased further (leading to higher angular momentum states becoming the ground state in the co-rotating frame), the number of vortices in the condensate grows which organize themselves in regular patterns \cite{br99,cd99}.
In the present work, we go beyond the slow rotating regime to focus specifically on moderately rotating regime with angular momenta $4N \le L_{z} < 5N$, well below the regime where the vortex lattice appears. 
For repulsive Bose-condensed gas rotating with angular velocity $\Omega < 1$, there exists a series of stable vortex patterns with discrete $p$-fold rotational symmetry, where the system is well described by a ground state with a definite vorticity $m_{1}$, the single-particle angular momentum quantum number corresponding to the largest condensate fraction $\lambda_{1}$ of the SPRDM~(\ref{spd}).
In order to gain an insight into the dynamics of formation of quantum mechanically stable vortical states with definite vortex patterns, we study stable as well as unstable states. 
The study reveals the path taken by the system to reach a stable state.
In particular, we examine the internal structure of the condensate by analyzing CPD~(\ref{cpd}) for stable as well as unstable states.
\begin{table}[!htb]
\caption{\label{tab1} For $N=16$ bosons in given subspaces of total angular momentum $4N \le L_{z} < 5N$, the lowest eigenenergy $E^{lab}_{0}$ (in units of $\hbar \omega_{\perp}$) of the states in the laboratory frame, the value of critical angular velocity $\Omega_{{\bf c}i}$, filling fraction $\nu$, with $p$-fold rotational symmetry of stable vortical states, the largest eigenvalue $\lambda_{1}$ and the corresponding single-particle quantum number $m_{1}$ of the SPRDM~(\ref{spd}). The results presented are calculated with interaction parameter $\mbox{g}_{2}=0.9151$ and range $\sigma=0.1$ of the repulsive Gaussian interaction potential~(\ref{gip}).}
\begin{ruledtabular}
\begin{tabular}{ccccccc}
$L_{z}$ & $E^{lab}_{0}(L_{z})$ & $\Omega_{\bf c}$ & $\nu$ & $p$ & $m_{1}$ & $\lambda_{1}$ \\ 
\hline 
64 & 104.73899 &&&& 1 & 0.3513 \\
65 & 105.71752 &&&& 1 & 0.3261 \\
66 & 106.66694 &&&& 6 & 0.4160 \\
67 & 107.67820 &&&& 1 & 0.3281 \\
68 & 108.70366 &&&& 1 & 0.3464 \\
69 & 109.66932 &&&& 1 & 0.2719 \\
70 & 110.60461 &&&& 7 & 0.4159 \\
71 & 111.58246 & 0.9835 & 2.66 & 5 & 6 & 0.4681 \\
72 & 112.60024 &&&& 7 & 0.2796 \\
73 & 113.59806 &&&& 7 & 0.4227 \\
74 & 114.62161 &&&& 7 & 0.3633 \\
75 & 115.58554 &&&& 6 & 0.3044 \\
76 & 116.51123 & 0.9857 & 2.28 & 6 & 7 & 0.5691 \\
77 & 117.56864 &&&& 6 & 0.3029 \\
78 & 118.56103 &&&& 7 & 0.4941 \\
79 & 119.55921 &&&& 7 & 0.5436 \\
\end{tabular}
\end{ruledtabular}
\end{table}
\\
\indent
For $N=16$ bosons in the moderately rotating regime with total angular momenta $4N \le L_{z} < 5N$, Table~\ref{tab1} lists values of the following: (i) lowest energy $E^{lab}_{0}\left(L_{z}\right)$ in the laboratory frame, (ii) the critical angular velocity $\Omega_{\bf c}$, (iii) filling fraction $\nu$, (iv) along with the respective $p$-fold rotational symmetry of the stable vortical states, (v) the largest eigenvalue $\lambda_{1}$ (vi) with corresponding single-particle quantum number $m_{1}$ of the SPRDM. 
From the table, we find that the angular momentum states $L_{z}=71$ and $L_{z}=76$ correspond to stable states. 
The remaining $L_{z}$-states are unstable and reveals the mechanism of entry, nucleation and pattern formation of vortices as the condensate goes from one stable vortical state to the other. 
To this end, we present in Fig.~\ref{fig:ltc} the CPD contour plots of stable as well as unstable states, depicting isosurface density profiles of rotating Bose-condensate.
The choice of reference point ${\bf r}_{0}=\left(x_{0},y_{0}\right)=\left(1.5,0\right)$ for the CPD plots corresponds to the line (the ray along the $+$ve $x$-axis in Fig.~\ref{fig:ltc}) along which there is spontaneous breakdown of symmetry. 
\paragraph*{Stable vortical states.} 
From the mechanical stability of the system, it follows that when the rotational angular velocity $\widetilde{\Omega}$ exceeds the confining frequency ${\omega}_{\perp}$ {\it i.e.} $\widetilde{\Omega}>\omega_{\perp}$, the center-of-mass of the Bose-condensate is destabilized \cite{rps02}.
When $\widetilde{\Omega}$ and $\omega_{\perp}$ are comparable {\it i.e.} $\widetilde{\Omega} \lesssim \omega_{\perp}$, the centrifugal force influences the shape of the condensate by strongly depleting the density along the axis of the trap \cite{fet09}, as shown in Fig.~\ref{fig:ltc}.
The central vortex emerges beyond the angular momentum states $L_{z} \ge 64$, in agreement with earlier mean-field results \cite{br99,cd99,kf04}.
It is to be noted that the vortex at the trap center is absent for stable vortical ground states with angular momentum $N < L_{z} < 4N$, whereas it is necessarily present both in stable as well as unstable states with angular momentum $4N \le L_{z} < 5N$ for $N=16$ bosons.
\\
\indent
The stable vortical state with $L_{z}=71$ corresponding to one of the critical angular velocity in our system of $N=16$ bosons, forms a pentagonal vortex pattern with a central vortex along the trap center \cite{br99,cd99} as seen in Fig.~\ref{fig:ltc71}. 
The vortex pattern with 5-fold rotational symmetry may exist in a finite, harmonically confined and hence inhomogeneous system being studied here \cite{kf04}, but will not form a translationally invariant infinite lattice.
From Table~\ref{tab1}, we also note that the stable vortical state with $L_{z}=71$ has vorticity $m_{1}=6$ with filling fraction $\nu=2.66$ and comprises of six vortices$-$five on the edges and one at the center of the trap.
It appears that each vortex corresponds to a singly quantized vortex carrying unit circulation.
Further, apart from the first stable vortical state $L_{z}=N=16$ aligned with the trap center, the central vortex reappears \cite{vvh05} only after $L_{z}=71$ stable vortical state with vorticity $m_{1}=6$, corresponding to the largest condensate fraction $\lambda_{1}$ of the SPRDM.
\\
\indent
The next stable vortical state seen in Table~\ref{tab1} is the total angular momentum $L_{z}=76$ state for $N=16$ bosons and is found to have vorticity $m_{1}=7$.
From the CPD plot shown in Fig.~\ref{fig:ltc76}, we observe that the vortical state $L_{z}=76$ possesses a 6-fold rotational symmetry forming a hexagonal vortex pattern with a central vortex along the trap center \cite{br99}.
The vortex pattern with vorticity $m_{1}=7$, thus, comprises of seven singly quantized vortices \cite{tap13} with filling fraction $\nu=2.28$, that is, one singly quantized vortex right at the trap center surrounded by six singly quantized vortices arranged on a hexagon \cite{vvh05,kf04,gbp11}.
In the limit of higher angular velocity ${\Omega}$, the centrifugal force significantly influences the shape of the condensate leading to nucleation of a vortex lattice. 
As discussed earlier, not all stable vortical states form a vortex lattice. 
The stable vortical state in Fig.~\ref{fig:ltc76} with 6-fold rotational symmetry may, at high angular velocity, form a lattice with regular triangular symmetry. 
It is important to mention that, though we do not clearly observe the triangular vortex lattice as suggested in \cite{coo08,fet09}, our exact diagonalization result on a finite system bears the signatures of vortex pattern with a central vortex at the trap center.
Transition between these stable vortex patterns are then studied by examining the internal structure (spatial correlation) of intervening unstable states.
\paragraph*{Unstable rotating states.}
In Table~\ref{tab1}, angular momentum states other than $L_{z}=71\ \mbox{and}\ 76$ (in angular momentum regime $4N \le L_{z} < 5N$ for $N=16$ bosons) are found to be unstable in the co-rotating frame.
The internal structure of these unstable states too exhibits patterns similar to stable vortical states but with less pronounced local minima of boson density, as seen in CPD contour plots of  Fig.~\ref{fig:ltc}.
We further observe that with regard to the number of local minima in density and its distribution around the center of the trap, the unstable states exhibit fluctuating behavior in the vicinity of stable vortical states. 
For instance, hexagonal patterns with $m_{1}=7$, appear for unstable states $L_{z}=70\mbox{ and }73$, shown in Figs.~\ref{fig:ltc70} and \ref{fig:ltc73} respectively, around the stable vortical state $L_{z}=71$ having a pentagonal vortex pattern, depicted in Fig.~\ref{fig:ltc71}, with $5$-fold rotational symmetry and vorticity $m_{1}=6$.
Similarly, pentagonal patterns with $m_{1}=6$ appear for unstable states $L_{z}=75\mbox{ and }77$, shown in Figs.~\ref{fig:ltc75} and \ref{fig:ltc77} respectively, around the stable vortical state $L_{z}=76$ having a hexagonal vortex pattern, depicted in Fig.~\ref{fig:ltc76}, with $6$-fold rotational symmetry and vorticity $m_{1}=7$. 
The unstable states may, thus, be viewed as carrying the imprints of the mechanism of pattern formation for stable vortical states, seen in Figs.~\ref{fig:ltc69} to \ref{fig:ltc78}.
Moreover, the series of CPD contour plots in Figs.~\ref{fig:ltc69}-\ref{fig:ltc78} also exhibits the melting of vortex patterns as a result of quantum fluctuation around the stable vortical states, due to uncertainty in positions of individual vortices.
Because of the harmonic confinement (resulting in inhomogeneous density), quantum fluctuation is most dominant at the trap center and consequently the melting of vortex patterns in unstable states, is most visible around and close to the trap center, wherever the density of bosons is nonzero.

\section{Summary and conclusion}
\label{conc}
The conditional probability distribution of stable as well as unstable angular momentum states in the co-rotating frame indeed reveals the mechanism of entry, nucleation and pattern formation of vortices as the BEC goes from one stable vortical state to the other with rotation.
We observe that after the first stable vortical state $L_{z}=N=16$ aligned with the trap center, the central vortex reappears only in the moderately rotating regime with $4N \le L_{z} < 5N$. 
In this regime, the stable vortical state $L_{z}=71$ has vorticity $m_{1}=6$ with filling fraction $\nu=2.66$ and comprises of six vortices$-$five on the edges arranged on a pentagon and one at the center of the trap.
This vortex pattern with five-fold rotational symmetry will not survive in the thermodynamic limit. 
The next stable vortical state $L_{z}=76$ in the regime considered above, possesses a 6-fold rotational symmetry forming a hexagonal vortex pattern with a central vortex along the trap center.
This six-fold vortex pattern is found to have vorticity $m_{1}=7$ with filling fraction $\nu=2.28$, and comprises of seven singly quantized vortices$-$one vortex right at the trap center surrounded by six vortices arranged on a hexagon.
Our exact diagonalization results on a finite system, thus, bear the signature of the thermodynamically stable triangular vortex lattice composed of singly quantized vortices.
The unstable states exhibit the melting of vortex patterns as a result of quantum fluctuation around the stable vortical states due to uncertainty in positions of individual vortices.

\end{document}